# Theoretical and observed potential energy curves for neutral 4-unit charge Coulomb systems containing antihydrogen

G. Van Hooydonk, Ghent University, Faculty of Sciences, Krijgslaan 281, B-9000 Ghent (Belgium)

Abstract. Comparing observed and theoretical potential energy curves for natural and exotic neutral 4-unit charge Coulomb systems like HH and HantiH leads to new conclusions on the effect of charge-antisymmetry in nature. With singularities in the HantiH PEC as found by Aldrovandi and Puget and by Junker and Bardsley, any cusp in the HantiH PEC significantly affects the annihilation cross section. This problem for the HantiH interaction generated many new wave mechanical calculations mainly to remove annoying cusps. We review all available PECs for 4-unit charge systems and find that corrections for the Morgan-Hughes HantiH PEC can either go to the *repulsive* side (to the conventionally *expected* annihilation channel) or to the opposite *attractive* side (to the attractive branch of the observed PEC of natural molecular $H_2$). We observe that all *theoretical* HantiH PECs published thus far would intersect the *observed* PEC of natural $H_2$. This is, however, impossible with the non-crossing rule. A classical *ab initio* calculation of the electrostatic perturbation brings the HantiH PEC much closer to the observed $H_2$ PEC than to the PEC of the annihilative channel. This new unprecedented result *for long-range behavior in 4-unit charge Coulomb systems* confirms that natural antiH must exist and that molecule $H_2$ must be interpreted as HantiH. In fact, this seems to be the only solution left, if PECs for systems HH and HantiH with the same symmetry cannot cross.

Pacs: 34.10.+x; 34.90.+q; 36.10.-k

*Introduction*

Understanding the stability of neutral 4 unit charge systems like prototype molecule $H_2$ has a long history, going back to the 19$^{th}$ century [1]. Heitler and London found the first wave mechanical solution for $H_2$ or HH [2] but we found evidence that molecule $H_2$ be denoted as H<u>H</u> [1]. This conclusion is based on observed short-range oscillator behavior in the PEC (potential energy curve) of natural $H_2$, which obeys the Kratzer potential around the minimum [1]. Yet, we need also a first principles explanation for long-range behavior near the asymptote [1] and must therefore prove that long-range behavior of $H_2$ is also consistent with the same *charge-anti-symmetrical* Hamiltonian of [1a]. To do so, we review all earlier wave mechanical calculations on H<u>H</u> [3], since we argued [1a] that the H<u>H</u> PEC was still uncertain, despite intensive work.

It has been conjectured for long that exotic 4-particle system H<u>H</u>, where neutral matter H interacts with neutral antimatter <u>H</u>, must annihilate. Recent claims on the mass production of artificial <u>H</u> [4] boosted theoretical studies on the H<u>H</u> PEC [3]. Yet there remains an uncertainty about the shape of the H<u>H</u> PEC as argued in [1]: exotic H<u>H</u> is *expected to be unstable because of annihilation* [3] but line and band spectra indicate that H<u>H</u> is *stable* instead [1]. We focus on this important controversy, since the work of Kolos et al. on H<u>H</u> [3a] and all more recent H<u>H</u>-studies [3b,c] were started mainly after *singularities* (energy extremes) were found earlier in the H<u>H</u> PEC [5-6]. With singularities, the *expected* annihilation of H<u>H</u> and its cross section is greatly affected [3,5-6]. About 30 years ago, the issue of extremes was important in the debate on the *best* shape of the H<u>H</u> PEC. Kolos et al. [3a] admitted the existence of a cusp but argued its height was far less than calculated earlier [5-6]. Since extremes in a H<u>H</u> PEC have far reaching consequences [5], not the least for the matter-antimatter asymmetry in the Universe [7], we review published H<u>H</u> PECs to find out what is at the roots of this dispute. Inventorying all PECs for neutral 4-unit charge Coulomb systems is an important classical exercise [1a] to understand these systems[1].

Today, the singularity problem in *theoretical* H<u>H</u> PECs *seems* solved satisfactorily, since it is believed generally that the H<u>H</u> attraction is continuous (*without extremes*) [3] so as to end finally in the *expected* annihilation of exotic bond H<u>H</u>. The implicit reason for this *expectation* is a *persistent belief* that, at short range, exotic H<u>H</u> *smoothly* rearranges in 2 neutral sub-systems *positronium* Ps and *protonium* Pn, which lead to annihilation [3,9]. *This expectation seems plausible*, since the 2 *charge-conjugated and mass-symmetrical* neutral 2-unit charge systems Ps and Pn will annihilate according to Dirac theory [1a]. But the perfect charge-symmetry in both Ps and Pn, needed for their annihilation to be possible is, in reality, broken by the large mass difference between lepton and baryon in H and <u>H</u> [10], both present in neutral 4 particle Coulomb aggregate H<u>H</u>. With a PEC

---

[1] some consequences of the existence of natural <u>H</u> [1] for the mass production of artificial <u>H</u> [4] are in [8]

representation, this is easily accounted for by assigning a different asymptote to Ps+Pn and H+H channels *at large inter-baryon separation* (see Fig. 4 in [3a] and further below). But *at short range*, all calculations assume –*without explicit proof*- that the PECs for the *charge-symmetrical channel* Ps,Pn and for *charge-anti-symmetrical channel* H,H will have to get close together, *in line with the expectation of the annihilation of the HH system*. Therefore, we must find out exactly what happens at intermediate range, where the curves for the two systems are getting closer, will or will not cross and where the chances for singularities to occur are real [5-6]. Whether or not HH PEC calculations are biased mainly by the *expectation of annihilation* is dealt with below, as, unfortunately, bias is not unusual in physics [11]. Quantum chemistry has proved unreliable in many instances too [12] as also argued in [1].

Theoretically probing singularities for *exotic* particle systems, *not yet observed experimentally*, is severely biased indeed. Since for a 4-unit charge system there is no disagreement at all on the 10 terms in the two Hamiltonians, the problem reduces to a number of arbitrary or subjective choices for the wave functions needed to solve the wave equation [1a]. First, there is the radial dependency of *atomic* wave functions STO and GTO. Their ability to deal with cusp conditions is critical as is their usefulness and reliability at long, intermediate and short range. Due to their analytical difference, STOs in general lead to a better convergence than GTOs, which do not give the correct lepton behavior near the nucleus. GTOs cannot satisfy cusp conditions and only additional terms can correct for *wrong* short-range behavior. For H-containing systems HH and HH, STOs seem the *best and natural* choice since they resemble the exact functions of H. Unlike GTOs, STOs exhibit the correct cusp behavior at short range, whereas GTOs are known to fall off too rapidly at long range. More terms are needed to achieve the same accuracy with GTOs than with STOs. But GTOs are nevertheless preferred over STOs because of reduced computing times, *not the most objective argument in terms of theory*. Recent work on the HH-system uses GTOs in the adiabatic Born-Oppenheimer approximation and predicts a very *smooth* HH PEC *without any singularity* [3c]. With charge-inversion, the nucleon-nucleon interaction changes suddenly from *repulsive* to *attractive*, creating an enormous Coulomb gap of $|2e^2/r_{AB}|$ between the 2 *mutually exclusive charge-conjugated states*, *charge-symmetrical* HH or HH and *charge-anti-symmetrical* HH or HH [1]. We try to cover most of these aspects in the analysis below after reporting extensively on the actual status of all published *theoretical* HH PECs.

*Inventory of theoretical HH PECs*

We collected all data on HH PECs from the original publications, published during the last decades, except those on the important HH PEC of Aldrovandi and Puget [5], for which no data were tabulated. From Fig. 6 in [5] and their discussion, we know they found a *minimum* at 2,8 a.u. below asymptote –1 and a *maximum* at 1,8 a.u. above this asymptote of +0,8 eV.

*With the exception of this unique HH PEC with its 2 extremes* [5], the shape of all theoretical HH PECs is illustrated in Fig. 1, where, as expected with annihilative HH, no singularity *seems* to occur at first sight. It is evident from Fig. 1 that the results of the primary HH PEC of Morgan and Hughes [9] (called δ in their paper) diverge markedly from those of Kolos et al. [3a], which, in turn, is almost indiscernible from the recent HH PECs of Labzowsky et al. [3b] and of Strasburger [3c]. A remarkable conclusion from Fig. 1 is that for Morgan and Hughes the HH interaction is much *more attractive at all ranges considered* than for all others. For instance, at R = 1,5 a.u., the difference between the 2 main PECs in Fig. 1 is ~30000 cm$^{-1}$, much too large[2] to be neglected, and which must be discussed properly.

In the global E, R view of Fig. 1, differences between HH PECs JB [6], KMSW [3a], LSPPS [3b], similar to S [3c], are hardly visible. One would be tempted to conclude from all these *theoretical* HH PECs, except for the Morgan-Hughes PEC also in Fig. 1, that the HH attraction is indeed well understood, since most of them converge towards the *theoretically expected* annihilation of neutral H and H, along the lines of the annihilation of the Ps+Pn channel (see Fig. 4 of [3a]). The only diverging *theoretical* HH PEC [9] is important enough to be examined more closely. In fact, Morgan and Hughes [9] took great care in obtaining a HH PEC just from the electrostatic perturbing part of the total Hamiltonian, an effort to be dealt with below.

---

[2] e.g. too large in comparison with the total well depth of the HH system at 0,74 Å of about 38200 cm$^{-1}$ (see below)

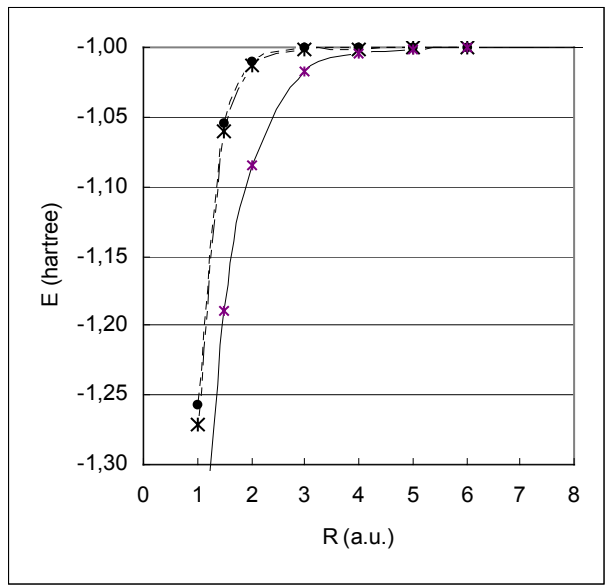

Fig. 1 Basic characteristics of main theoretical HH PECs
(full line with x: primary HH PEC called δ by Morgan and Hughes [9], dashed lines with ●, * all (nearly coinciding) other HH PECs: Junker-Bardsley [6], Kolos, Morgan, Schrader, Wolniewicz [3a], Labzowsky, Shapirov, Prozorov, Plunien, Soff [3b], Strasburger [3c])

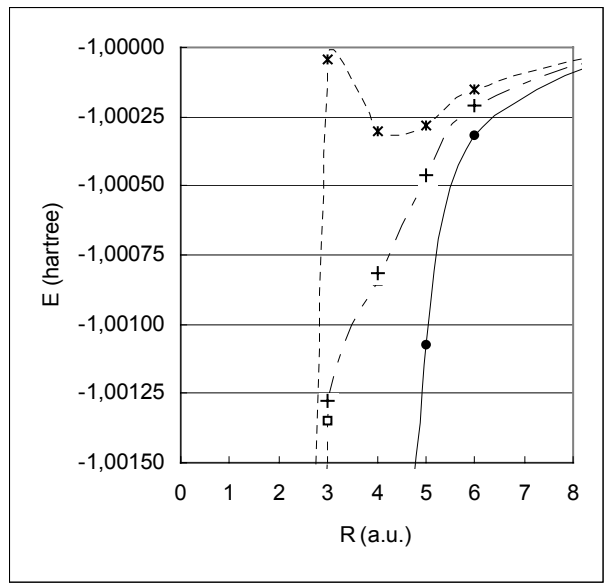

Fig. 2 Singularities in theoretical HH PECs
(full line with ●: primary HH PEC called δ by Morgan and Hughes [9], dashed lines with * HH PEC of Junker-Bardsley [6], with + Kolos, Wolniewicz, Morgan, Schrader [3a], with □ Strasburger [3c])

Apart from the 2 extremes in the Aldrovandi-Puget HH PEC [5], Fig. 2 gives the details of the cusp below the asymptote found by Junker and Bardsley [6]. We adapted the scales to illustrate the JB-barrier around $3r_0$ (~1.5 Å) in comparison with other *theoretical* PECs, where this extreme is much smaller [3a] (for similar figures, see Fig. 1 in [6] and Fig. 1 in [3a]).

Fig. 1 and 2 illustrate part of the difficulties with *theoretical* HH PECs, mentioned in [1a]. Both are constructed with data, published by a variety of scientists working on the *theoretically expected* PEC for so-called exotic 4-unit charge Coulomb system HH. They set out the further contours for our analysis. With Fig. 2, it is impossible to conclude like in [3c], that the HH interaction *is already relatively well known* and that the H<u>H</u> PEC *has been established for over 20 years* [3c]. On the contrary, Fig. 2 shows that, in reality, *theoretical* H<u>H</u> PECs diverge markedly *at all ranges of interest: long, intermediate as well as short*. The differences are not small but huge (see also footnote 1 and Appendix), which does not simplify the situation, since, in addition, the experimental H<u>H</u> PEC has not yet been measured (but see below). In the absence of this PEC, a discussion of the relative merits and precision of all above theoretical H<u>H</u> PECs is academic. One conclusion is certain: some H<u>H</u> calculations will have to be revised since they cannot all be correct at the same time. The problem is now to find out which is (are) wrong and why.

*Crossing of theoretical HH PECs with the only observed PEC for a stable 4-unit charge Coulomb system*

Obviously, the absolute reference PEC missing in Fig. 1 and Fig. 2 is the PEC of the only natural, neutral and stable 4-unit charge system $H_2$, known with sufficient spectroscopic accuracy for many a decade [13]. It is therefore of ultimate interest to compare the theoretical PEC of the so-called exotic H<u>H</u> system with that of HH or $H_2$, since the two are both neutral 4-unit charge Coulomb systems with the same symmetry. This exercise, never done properly when it was most needed, it is not without interest [1a]. In fact, the remarkable result of bringing all PECs together is shown in Fig. 3.

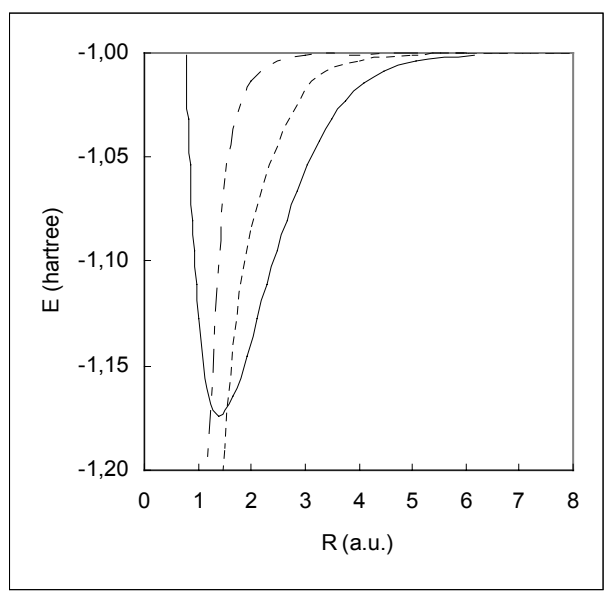

Fig. 3 Theoretical H<u>H</u> PECs and the observed PEC of $H_2$
(full line observed $H_2$ PEC [13], short dashes Morgan-Hughes H<u>H</u> PEC [9], mixed dashes H<u>H</u> PEC Kolos et al. [3a]

It is strange that the *observed* $H_2$ PC in Fig. 3, published in 1963 [13], was *theoretically reproduced* with great accuracy by Kolos and Wolniewicz in 1965 [14] (see also Appendix) and that the authors' great experience with the $H_2$ PEC was used to calculate, together with Morgan and Schrader [3a], the *theoretical* H<u>H</u> PEC, also shown in Fig. 3. It is now evident that a primary objective of [3,9] must have been to remove the unwanted singularities detected in earlier H<u>H</u> PECs (see Fig. 2) in order to make sure that the *expected annihilation of HH would indeed proceed as generally believed*.

*The most important and imminent conclusion from Fig. 3 is that it simply contradicts the conventional belief that H<u>H</u> will annihilate. Since, in wave mechanics, HH and H<u>H</u> have the same symmetry, their PECs are not allowed to cross. This definitely settles our point in* [1] *on the so-called annihilation of system H<u>H</u>, which is proved at least doubtful, if not impossible, with Fig. 3.*

It is therefore difficult to understand why all theorists did their very best to shift the primary Morgan-Hughes HH or δ PEC in Fig. 1 and 2 to the left (*to the more repulsive side*) where it would be closer to the PEC of the annihilative channel Ps+Pn, if they had known the position of this function with respect to the $H_2$ PEC (see Fig. 3). Even Morgan and Hughes [9] had no less than 3 suggestions to shift their own primary δ-function to the left with the adapted functions (PECs), they called α, β and γ (see Fig. 2 in [9]). For reasons not easily understood, no one ever tried to shift this same Morgan-Hughes δ PEC to the right (*to the more attractive side*), where it would have come close to the attractive branch of the observed $H_2$ PEC, a plausible possibility when looking at Fig. 3 and the crossing problem therein. These *theoretical* HH PECs, published a long time ago, were in fact one of the main reasons for our fundamental dissatisfaction with wave mechanical procedures [1,12] and with their reliability and *predictive power for not yet observed systems* [1a] (see below). Using Fig. 3, it is obvious that these calculations, all giving rather smooth HH PECs, *intentionally corrected the perturbing Coulomb terms of the Morgan-Hughes δ-function for HH* towards the annihilative channel *at the left*, whereas the observed attractive branch of the $H_2$ PEC was *the only known and more reliable reference channel* for correcting this same HH function *to the right* [1] with the prospect of avoiding crossings.

Since in [1a] we concentrated on the minimum of the $H_2$ PEC, we now easily verify with Fig. 3 that, at the minimum of $H_2$, all theoretical HH PECs, without exception, would have to cross the observed $H_2$ PEC. *This evident possibility is the more remarkable since anyone could have constructed Fig. 3 many decades ago, all data being available in the 1970s. If this exercise had been done in due time* [1a], *one would have been confronted at a very early stage with the same surprising situation: the crossing of theoretical annihilative HH PECs with the observed HH PEC at the minimum* and with all of the drastic consequences thereof. In any case, Fig. 3 clearly reveals that something very fundamental may have been overlooked indeed in all efforts above to predict a *theoretical* PEC for 4-unit charge Coulomb system HH [1a]. For if the theoretical HH PECs in Fig. 3 were really reliable as claimed by the authors, the non-crossing rule would secure that, at the crossing point, all annihilative HH PECs would invariantly go over in the *repulsive, non-annihilative* branch of the observed HH or $H_2$ PEC. Similarly, the non-crossing rule would secure that the observed *non-annihilative* $H_2$ PEC would go over in the *annihilative* HH PEC, which is impossible as $H_2$ is stable. Since of all PECs in Fig. 3 only the *observed* HH or $H_2$ PEC is really reliable, the non-crossing rule provides with an almost absolute criterion for shifting all of the theoretical HH PECs to the right, the only direction possible to coincide with the attractive branch of HH or $H_2$ like we argued in [1].

To solve this problem, we must be able to account for long-range behavior in 4-unit charge systems using the same Hamiltonian approach of [1a]. Oscillator behavior around the minimum in the $H_2$ PEC in Fig. 3 is sufficiently understood with the Kratzer potential hidden in the 10-term attractive Hamiltonian [1a]. At long range, oscillator behavior smoothly transforms in non-oscillator behavior, a transition left unaccounted for in [1a] (see the left part of Fig. 4b in [1a]). To remedy for non-oscillator behavior with an *ab initio* treatment, we are obliged to test this very same bound state Hamiltonian, *with its built-in Kratzer potential at short range* [1a], for its prediction at long-range for 4-unit charge Coulomb systems. Looking at Fig. 3, this is indeed a crucial and critical problem for all 4-unit charge PECs reviewed here.

*Long range behavior of charge-symmetrical and charge-anti-symmetrical 4-unit charge systems and their PECs*

Natural HH and its symmetrical *exotic* counterpart HH obey identical Hamiltonians, since a double intra-atomic charge inversion leaves the total 10-term Hamiltonian invariant [1a,3a]. The single charge-inversion needed for charge-antisymmetric aggregates HH or HH alters the sign of the 4 inter-atomic Coulomb terms in their Hamiltonians, which makes these *charge-antisymmetric (charge-conjugated)* too without interference of lepton or baryon spin. With [1], the combined 4-unit charge algebraic Hamiltonians

$$\mathbf{H}_\pm = \mathbf{H}_0 \pm \Delta\mathbf{H} \tag{1}$$

are, by definition, mutually exclusive since they are *(charge-)conjugated*, despite the neutrality of all 4-unit charge Coulomb species involved [1a].

The attribution of the signs in (1) is arbitrary, since it is impossible to solve 4 particle systems classically, unless geometrical models[3] are imposed. Hamiltonians (1) can safely be regarded as absolutely valid, pending the geometric 4-unit charge model. In the adiabatic Born-

---

[3] Using geometric models (leading to form factors) is the classical equivalent of using wave functions

Oppenheimer approximation of wave mechanics, a few uncertainties and arbitrary criteria are introduced. The first uncertainty must reside in the wave functions used for solving the wave equation [1] as remarked in the Introduction (see Appendix for more details). In [1a], we argued that putting the value of the wave equation equal to +1 in whatever region concerned focuses the attention to the algebra in total Hamiltonian (1) and, especially at longer range, to the electrostatic perturbing terms in $\pm\Delta\mathbf{H}$. As in [1], we use classical geometrical models to mimic the effect of wave functions in order to conserve the transparency required for a classical discussion of the four perturbing Coulomb terms in (1).

With Hamiltonians (1), the next problem is to find out how the very same 4-particle Hamiltonian for the bound state, which hides a Kratzer potential at the minimum [1], can also come to the rescue at longer range (attractive branch of the PEC) for the very same system when the Kratzer potential is no longer obeyed [1].

In addition, to avoid ambiguity and bias, we must secure that the evidence for a shift to the right of the Morgan-Hughes H$\underline{H}$ PEC in Fig. 1-3 has *ab initio* status before it can become competitive with the conventional shift to the left. As stated above, Morgan and Hughes spent much time on the perturbing terms in the H$\underline{H}$ Hamiltonian and wrote: 'However, for all values of (the inter-baryon separation) $r_{AB}$, it is probable that V (H$\underline{H}$ interaction energy) is negative. For $r_{AB}>r_0$ (Bohr radius), V is *roughly the negative of V for the corresponding atom-atom pair*' [9]. These statements are of *ab initio* status, since they are a direct consequence of the conjugation in (1) [1a]. For relatively simple 4-unit charge Coulomb hydrogenic systems of comparable complexity like HH and H$\underline{H}$, Fig. 1-3 prove that wave mechanical calculations do not agree [1a].

Using the approximation +1 for the wave function [1a], it is simple to verify what the 4-unit charge Hamiltonian itself says about the H$\underline{H}$ interaction for a given configuration. This procedure is particularly suited, given the efforts by Morgan-Hughes on the four perturbing terms in a wave mechanical procedure without a specific model [9].

We first scale (1) with the Hartree $|e^2/r_0|$ ($r_0 = 1$ a.u.) and filter out the (charge-) conjugated perturbing terms [1b], which gives

$$\pm\Delta\mathbf{H}/(|e^2/r_0|) = \pm r_0 (-1/r_{Ab} - 1/r_{Ba} + 1/r_{ab} + 1/r_{AB}) \quad (2)$$

where subscripts a and b refer to the leptons, A and B to the baryons [1a]. The + solution of (2) gives the HL-perturbation for *charge-symmetrical* $H_2$ (atom-atom system), the – solution applies for the *charge-anti-symmetrical* perturbation in H$\underline{H}$ (atom-anti-atom system) both in the ground $\Sigma$-state (spin must not be considered [1a,3a]).

To solve electrostatic problem (2), we use a configuration, *where particle rearrangements are not allowed* to preserve transparency. To keep the system dynamic also, we use Bohr-like molecular models [15]. The leptons are both in orbit in planes perpendicular to the inter-baryon separation $r_{AB}$ and each of them describes a circular motion with radius $r_0$, the Bohr length. Despite its simplicity, this model without polarization [1b] allows calculations of *ab initio* or first principles' status with the added advantage to discuss deviations from the model with *classical physics*. Revival in the interest for Bohr's molecular models proves their efficacy [16] and confirms the efficiency, reliability and transparency of old quantum theory in comparison with wave mechanics [1].

With this Bohr-model, atom-atom case HH gives parallel dipoles or ↑······($r_{AB}$)······↑, if lepton motion is synchronous and without phase-shift [1b,17]. Similarly, atom-anti-atom case H$\underline{H}$ uses anti-parallel dipoles ↑······($r_{AB}$)······↓ with the same constraints for lepton motion (other cases are discussed elsewhere [17]).

For very large separation, the lepton-lepton and baryon-baryon separations obey

$$R= r_{ab}=r_{AB} \quad (3)$$

which enforces a simple classical solution also for lepton-baryon separations

$$r_{Ab} = r_{Ba} = \sqrt{(R^2+r_0^2)} \quad (4)$$

Using the standard identity to express separations in a.u. or $r_0$

$$R = mr_0 \quad (5)$$

which secures $r_0/R= 1/m$.

With (3)-(5), the perturbing Coulomb terms (2) simply reduce to [1b]

$$\pm\Delta\mathbf{H}/(|e^2/r_0|) = \pm (2/m)(1- 1/\sqrt{[1+(1/m^2)]}) \quad (6)$$

This is an equation with model-dependent *ab initio* status and provides with the transparency missing in wave mechanical procedures [1a]. We remark that Morgan and Hughes solved perturbation (2) with wave functions tested for *good behavior* with solving the wave equation for H<u>H</u> [9]. This is a typical case where wave mechanics looses the transparency of classical physics.

For large m, dipole-dipole interaction (6) simplifies to *repulsive* $+1/m^3$ for HH and to *attractive* $-1/m^3$ for its conjugated or charge-anti-symmetrical counterpart H<u>H</u> [1b]. For obvious reasons, we will not plot $+1/m^3+...$ for the repulsive interaction in *charge-symmetrical* HH, since repulsion can never lead to annihilation [1] and since $+1/m^3$ is the conjugate of attractive $-1/m^3$ for H<u>H</u>, which we will use below. Of course, at very long range (very large m), it is difficult to assume that the 2 main classical constraints (synchronism of phase-less lepton motion) are applicable. Attenuation of the underlying pure Coulomb interactions in (6) will have to be introduced to accommodate for *less ordered* states (e.g. asynchronous motion with phase-shifts) with statistical correction factors, say of exponential type [17], with dispersion forces[4], polarization effects…). Most of these, as well as similar effects are discussed in [17].

*Results and discussion*

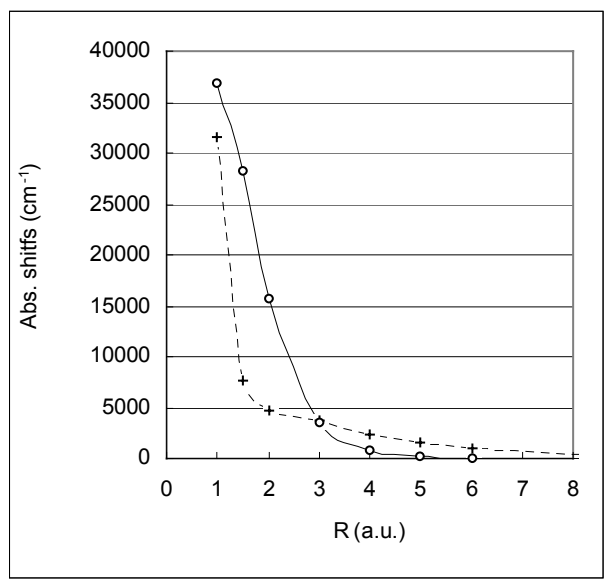

Fig. 4 Absolute shifts (in cm$^{-1}$) for the Morgan-Hughes δ-function for H<u>H</u> [9]: to the right with this work, equation (6) (dashed line) and to the left with Strasburger's H<u>H</u> PEC [3c] (full line)

First, we compare the magnitudes of the shifts to the repulsive left, implicit in the H<u>H</u> PEC of Strasburger [3c] and to the attractive right, based on (6). With the Morgan-Hughes PEC as reference, the theoretical right-shift with (6) is *less drastic* than the left-shift suggested in [3c] as illustrated in Fig. 4. We are confident in the *theoretical* reliability of (6), since Morgan and Hughes devoted so much attention [9] to perturbation (2).

Next, despite the obvious but transparent constraints of our dipole-dipole model, we have plotted the attractive –solution of (6) also in Fig. 5 to show how the long range PEC deriving from (6) behaves in comparison with the PECs in Fig. 3. The classical interaction energy for H<u>H</u> with (6) shifts the Morgan-Hughes PEC indeed to the right, i.e. to the more attractive domain, instead of to the left in the annihilative model. We must now decide which shift is the better to remove the *wrong behavior* of the primary Morgan-Hughes H<u>H</u> or δ PEC in Fig. 3 and to avoid the (not allowed) crossing of the HH and H<u>H</u> PECs. The conventional shift to the left is clearly biased by the universal *expectation* of HH̲ annihilation (rearrangement to Ps+Pn in an annihilative channel). On the contrary, *a theoretically validated shift to the right*, due to (6), brings the right-corrected HH̲ PEC close to the attractive branch of the *observed* $H_2$ PEC. This seems a strange result but it is nevertheless also the transparent result of a classical *ab initio* calculation (6), the validity of which cannot be denied despite its 2 straightforward constraints on lepton motion.

---

[4] Both the primary Morgan-Hughes PEC and ours are much more attractive than Roothaan's electrostatic interaction $E_{es} = (-1/R - 5/8 + 3R/4 + R^2/6)e^{-2R}$, which at R=10a.u. gives +0,04832 μHartree. Morgan-Hughes give –7,96 μHartree and we obtain with (6) –0,99 mHartree. A similar remark applies for dispersion forces, varying as $-C_6/R^6 - C_8/R^8 - ...$ but not discussed explicitly here (for details see [17]).

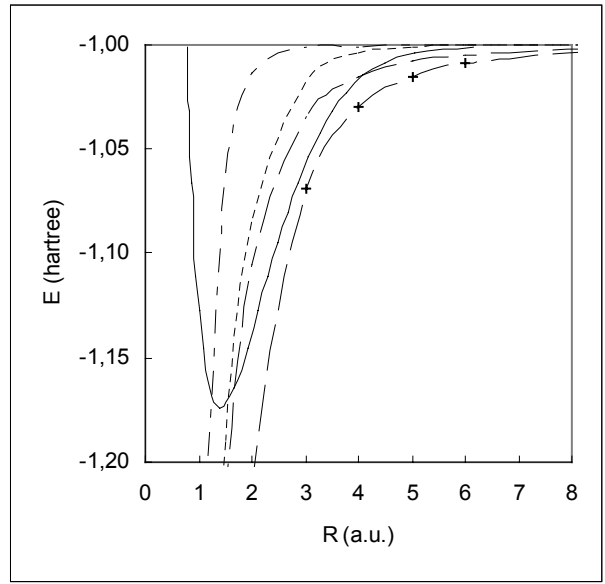

Fig. 5 Confrontation of HH̲ PECs in this work with the theoretical and observed PECs in Fig. 3: long dashes, equation (6), long dashes with +, Coulomb HH/HH̲ gap, equation (7)

It is now likely that all expectations on the annihilation of HH̲ will have to be reviewed. One can simply verify *de visu* that of all theoretical HH̲ PECs collected above, only long range attraction obeying (6), i.e. the long range HH̲ PEC with our model, is closer to the attractive branch of the observed PEC of $H_2$ and would be capable to avoid the crossing problem in Fig. 3. This is another unprecedented result, adding to those reported earlier on the PEC of a bound 4-unit charge system around the minimum [1]. More details on the effects of *antisymmetry* are in [17].

Furthermore, and as shown in [1b], the scaled Coulomb gap C between states HH and HH̲ with (6) is

$$C = -(4/m)(1 - 1/\sqrt{[1+(1/m^2)]}) \quad (7)$$

and is expected to be valid at long range, near the dissociation threshold. The corresponding PEC is also given in Fig. 5. However, the simple linear relation between (7) and observed long-range behavior of $H_2$ reported in [1b] opens the way for natural H-H̲ oscillations. Seen in retrospect, it is remarkable that theoretical work on the HH̲ PEC was intensified with the appearance of the *small* cusp in Fig. 2 but that the *large* cusp in Fig. 3 was not considered when trying to construct a PEC for neutral 4-unit charge Coulomb systems like HH and HH̲.

With Fig. 5 and Fig. 4a in [1a], we managed to understand satisfactorily the complete $H_2$ PEC using the so-called *wrong* Hamiltonian, believed to apply only for exotic system HH̲. This is likely to remain an extraordinary result for Physics/Einstein Year 2005 [1a]. In short, (i) molecule hydrogen $H_2$ should be understood with charge-anti-symmetrical HH̲ rather than with charge-symmetrical HH and (ii) *matter* (H) *does not annihilate when it meets antimatter* (H̲). Instead, when brought together, they form a very stable molecule $H_2$ [1].

*Conclusion*

A direct confrontation of all (theoretical and observed) PECs for neutral 4-unit charge systems reveals that *theorists* may well have been wrong footed by nature on how the interpret molecular hydrogen $H_2$. The predictive power of wave mechanics is indeed overestimated for not yet observed systems, where its predictions should be regarded with great caution [1a]. In fine,

Fig. 3 provides with a serious test for the reliability and transparency of wave mechanics in comparison with (Bohr's) old quantum theory [1a,16].

Despite conviction, $H_2$ is a stable non-annihilating matter-antimatter aggregate H<u>H</u> and it will require some courage indeed to admit that something very elementarily went wrong very early [1]. In fact, a straightforward way out of this dilemma with crossing in Fig. 3 is to abandon the idea that H and <u>H</u> will annihilate. If confirmed, experimental physicists and chemists can, fortunately, carry on as usual with *observing* PECs but for theorists, the situation is completely different. Finally, the so-called matter-antimatter asymmetry of the Universe [5,7] is no longer a problem for classical stochiometric reasons [1b].

**Appendix**
Uncertainties, arbitrary criteria and bias to solve the wave equation for 4-unit charge system HH̲

(i) First arbitrary criterion: lepton and nucleon separations

Unlike the HL treatment of $H_2$ [2], it soon appeared that, in order to solve the wave equation for HH̲, the variables had to be adapted, e.g. that the inter-lepton separation $r_{ab}$ was also required. Although it appears only in the perturbing parts $\Delta\mathbf{H}$ of total Hamiltonians (1), lepton separation $r_{ab}$ is indeed needed, following the Hylleraas solution for neutral three unit charge system He [19] and the James-Coolidge solution for neutral four unit charge system $H_2$ [20]. Since electron correlation was taken into account, this last solution for $H_2$ was significantly better than original Heitler and London theory [2]. But the introduction of $r_{ab}$ necessitated the use of more complicated wave functions, adapted for this famous Coulomb problem $+e^2/r_{ab}$. This increasing complexity brings in a first arbitrary criterion to decide on the *goodness* of a wave function. This goodness is usually only established *a posteriori*, i.e. after a calculation due to the mediocre transparency of wave mechanical procedures for many particle systems.

With (1), the lepton interaction for *charge-symmetrical* systems HH and H̲H̲ is *repulsive* $+e^2/r_{ab}$ (Coulomb problem) but for *charge-anti-symmetrical* HH̲ or H̲H this perturbing term becomes *attractive* $-e^2/r_{ab}$. This would remove the (lepton) Coulomb problem at once, if it ever existed [1a]. Whether or not a singularity occurs in the PEC for HH̲ is closely related with the analytical treatment of this famous Coulomb term for HH̲ or $H_2$. A similar conjugation also applies to the inter-baryon interaction, which is repulsive $+e^2/r_{AB}$ for *charge-symmetrical* systems but attractive $-e^2/r_{AB}$ for *charge-anti-symmetrical* systems. This huge charge-based parity effect has a tremendous impact on *the adiabatic Born-Oppenheimer approximation* in the first place.

(ii) Second arbitrary criterion: pathological behavior and the prospect of annihilation

In addition, it is known for long in a variety of disciplines that PEC-generating functions can easily result in *pathological behavior* [21], whereby parts of the PEC do not behave properly or *as expected*, a second arbitrary, if not subjective, criterion to judge on a given function. *In short, one expects some kind of PEC from the start and, therefore, one will intuitively select only these functions, which will obey these (subjective) expectations.* As pointed out earlier, this is the reason why the challenge for the reliability of wave mechanics is *to predict exactly, and theoretically from first principles only, the real PEC for an exotic particle system, for which the experimental PEC is not yet available* [1a]. Forms of pathological behavior are not restricted to theoretical PECs for *exotic* systems. In fact, they are a real bias for quantum chemical calculations at large [12], in agreement with our earlier assessment [1a].

The annihilation criterion is the main cause why the original and very elaborate M-H treatment [9] of the perturbing terms in (1) was abandoned and why, in the end, the M-H PEC for HH̲ in Fig. 1 *had to be shifted* to the more repulsive side (see Fig. 3). Only in this way, the PEC could get closer to the annihilation of the Ps+Pn channel, which is *believed* –without absolute proof given- to be the best channel also for the *expected* annihilation of HH̲.

(iii) Third arbitrary criterion: STO versus GTO, parameterization

After changing the variables to remedy for the Coulomb problem, another difficult problem is to select a *properly behaving wave function*. Here, the choice between STOs and GTOs is very important for the reasons given in the Introduction. Differences between the STO treatment of Kolos et al. [3a] and the GTO analysis of Strasburger [3c], *both believed to be very accurate*, are illustrated in Fig. 6. It is evident that something must be wrong even here, although the scales used in conventional Fig. 1 prevent the exposure of these discrepancies.

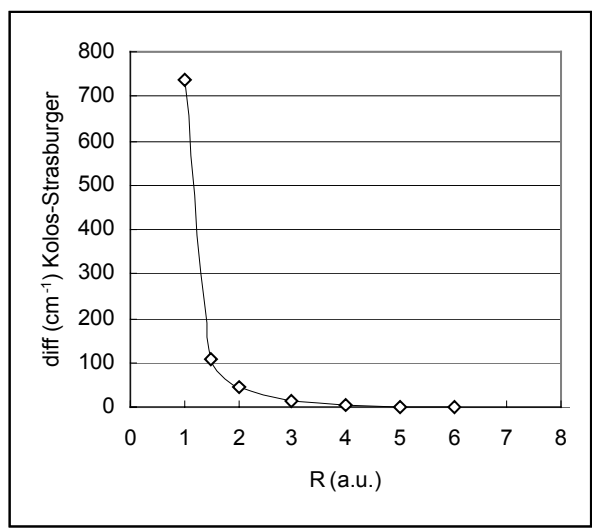

Fig. 6 Differences (in cm-1) between the 2 *best* HH̲ PECs: Kolos et al. [3a] (STO) and Strasburger [3c] (GTO)

The Strasbuger PEC for HH̲ [3c] is more attractive than that of Kolos et al. [3a]. To decide which of the 2 PECs is the better, one can only rely on the arguments put forward by the authors, since there is no external reference point. Reminding that the $H_2$ PEC, calculated by Kolos and Wolniewicz [14] is claimed accurate far below the 1 cm$^{-1}$ level (or far below µHartree), the differences between the 2 theoretical H$\underline{H}$ PECs are huge. In addition, the authors themselves warn about the relative accuracy of their PECs. Kolos et al. [3a] say that the accuracy of their H$\underline{H}$ PEC deteriorates quickly with decreasing $r_{AB}$. At $r_{AB}$ =3, the interaction energy converges to 6 significant figures but at shorter separation, *no more than 4 figures are reliable*. Even then, the uncertainty with Fig. 6 remains: it is simply impossible to decide which of the 2 functions is the better. One can only wonder why the $H_2$ PEC of Kolos and Wolniewicz [14] converges to 7 or 8 significant figures (nanoHartree) and their H$\underline{H}$ PEC to only 6 or even 4, when HH and H$\underline{H}$ are 2 very similar systems in terms of complexity, as remarked in [1a].

(iv) Arbitary procedures: a few explicit examples

It now is easier to understand a variety of remarks by authors, when speaking of a number of criteria *in order to secure proper behavior of the H$\underline{H}$ PEC, generated with the aid of particular wave functions.*

(a) For Kolos et al. [3a] there are problems for different regions of $r_{AB}$. "One set of 42 additional functions was used for $r_{AB} \geq 3$ a.u. and another for $r_{AB} < 3$ a.u.". This is exactly the region where also the accuracy of the HH PEC is greatly affected (see foregoing paragraph).

(b) Armour, Carr and Zeman [22] even introduce an extra function $\Psi_{ps}$ to represent *virtual positronium*. This function depends on $r_{ab}$, with a variational parameter $\varkappa$ as well as a shielding function $g(\varrho)$ to ensure *good behavior* of $\Psi_{ps}$ at $\varrho=0$. For this shielding function, depending on $(1-e^{-\gamma\varrho})^n$, to be workable, they write: "After some numerical trials, n was taken to be 3 and γ to be 0,5" … without any further explanation. They confirm that the H$\underline{H}$ interaction is smooth and without singularity but restrict their H$\underline{H}$ calculations to very short range (≤1). This is a range of less interest, where large errors are unavoidable (see the slopes in Fig. 1-3) and where energies exceed the total energy of $H_2$.

(c) Campeanu and Beu [23] say that the cusp in the Junker-Bardsley PEC (see Fig. 2) is *spurious* and that the PEC of Kolos et al. [3a] is better.

(d) Strasburger [3c] admits that there are 2 interesting *questions* with H-$\underline{H}$ in the BOA. The first is, as usual, the existence of a barrier in the interaction (see Fig. 2). He writes: "It is not easy to state definitely the shape of the PEC in the system, where both atom and anti-atom are neutral" (sic)…and continues with "The problem (of the barrier) was solved in the paper of Kolos et al.[3a] – the interaction appeared to be monotonic and attractive in the whole range of $r_{AB}$". His second question is related to the binding of light particles by baryonic matter… and he uses arguments provided by the strange procedure in [22], mentioned under (b) and is also mainly interested in providing more accurate values at short range (see Fig. 6). This concern with short range behavior is, again, an example of how bias due to the annihilative Ps+Pn channel enters the scene.

These examples illustrate what we meant by saying that *…in some of the works on H$\underline{H}$, rather subjective criteria are used for selecting so-called well-performing or well-behaving wave functions* [1a]. It appears we are indeed entitled to use this qualification, since the criteria in use are always governed by the conventional bias of how the H$\underline{H}$ PEC should look like *intuitively* or behave *properly* (*non pathological*), i.e. it should behave just like the annihilating Ps+Pn channel at short range.